\documentclass[journal]{IEEEtran}

\usepackage[utf8]{inputenc}
\usepackage[T1]{fontenc}
\usepackage{graphicx}
\usepackage{amsmath,amssymb,amsfonts}
\usepackage{textcomp}
\usepackage{xcolor}
\usepackage{booktabs}
\usepackage{multirow}
\usepackage{algorithm}
\usepackage{algorithmicx}
\usepackage{algpseudocode}
\usepackage{upgreek}
\usepackage{cite}

\begin{document}

\title{Active Lubrication of Transluminal Medical Instruments}

\author{Mostafa A. Atalla$^{1,2,*}$, Jelte Nieuwenhuis$^{1}$, Alan Martin$^{3}$, Xuan Wang$^{3}$, Ahranee Canden$^{3}$,\\Matt J. Carr\'{e}$^{3}$, Roger Lewis$^{3}$, Aim\'{e}e Sakes$^{1,\dagger}$, and Micha\"{e}l Wiertlewski$^{2,\dagger}$%
\thanks{$^{1}$ Department of BioMechanical Engineering, TU Delft, The Netherlands.}%
\thanks{$^{2}$ Department of Cognitive Robotics, TU Delft, The Netherlands.}%
\thanks{$^{3}$ School of Mechanical, Aerospace and Civil Engineering, University of Sheffield, United Kingdom.}%
\thanks{$\dagger$ A. Sakes and M. Wiertlewski are co-senior authors.}%
\thanks{* Corresponding author: Mostafa A. Atalla (e-mail: m.a.a.atalla@tudelft.nl).}}

\maketitle

\begin{abstract}
Transluminal minimally invasive surgery uses natural orifices and small incisions to access internal anatomical structures, promoting quicker recovery and reduced morbidity. However, navigating instruments--catheters and endoscopes--through anatomical pathways creates frictional interactions with luminal walls, risking complications such as perforation, poor haptic feedback, and instrument buckling. This paper presents an active lubrication sheath that controls friction on demand through discrete friction control modules distributed along its shaft. These modules employ ultrasonic vibrations at the instrument surface to generate a pressurized fluid layer at the contact interface, lubricating the interface and thereby reducing friction. We implemented these modules in a prototype catheter, which we validated under dry and liquid-lubricated conditions, across rigid and soft interfaces, and along varied anatomical curvatures. In a cardiac catheterization use case, active lubrication reduced friction by up to $42\%$ on \textit{ex vivo} porcine aorta tissue hydrated with phosphate-buffered saline, and thermal imaging measured an average temperature rise of $6.2^{\circ}\mathrm{C}$ at the module-tissue interface due to vibration, confirming its potential thermal safety. In a catheter insertion demonstration, active lubrication reduced friction-induced buckling and enabled smoother advancement through the lumen, further showcasing its potential impact. By minimizing injury risk and enhancing procedural stability, active lubrication can drastically enhance the safety and efficacy of transluminal interventions.

\end{abstract}


\section{Introduction}
Minimally invasive surgeries have reshaped healthcare by reducing patient trauma, shortening recovery time, and improving clinical outcomes. Among these techniques, transluminal interventions, which access internal anatomy through natural orifices and small incisions, enable procedures at difficult-to-reach anatomical targets~\cite{wagh2007,Omisore2022,reviewSoroMIS}. These interventions typically follow a two-phase structure: first, instruments navigate through anatomical pathways to the target, and second, once the instrument is positioned, surgical tasks are performed at the target site. Because tissue interaction differs markedly between these phases, the design of transluminal instruments must accommodate contrasting functional requirements. In particular, friction must be low during insertion for smooth navigation and minimal tissue damage, yet needs to remain high when the instrument is at the target location for stability. 
 
\begin{figure*}
    \centering  \includegraphics[width=\textwidth]{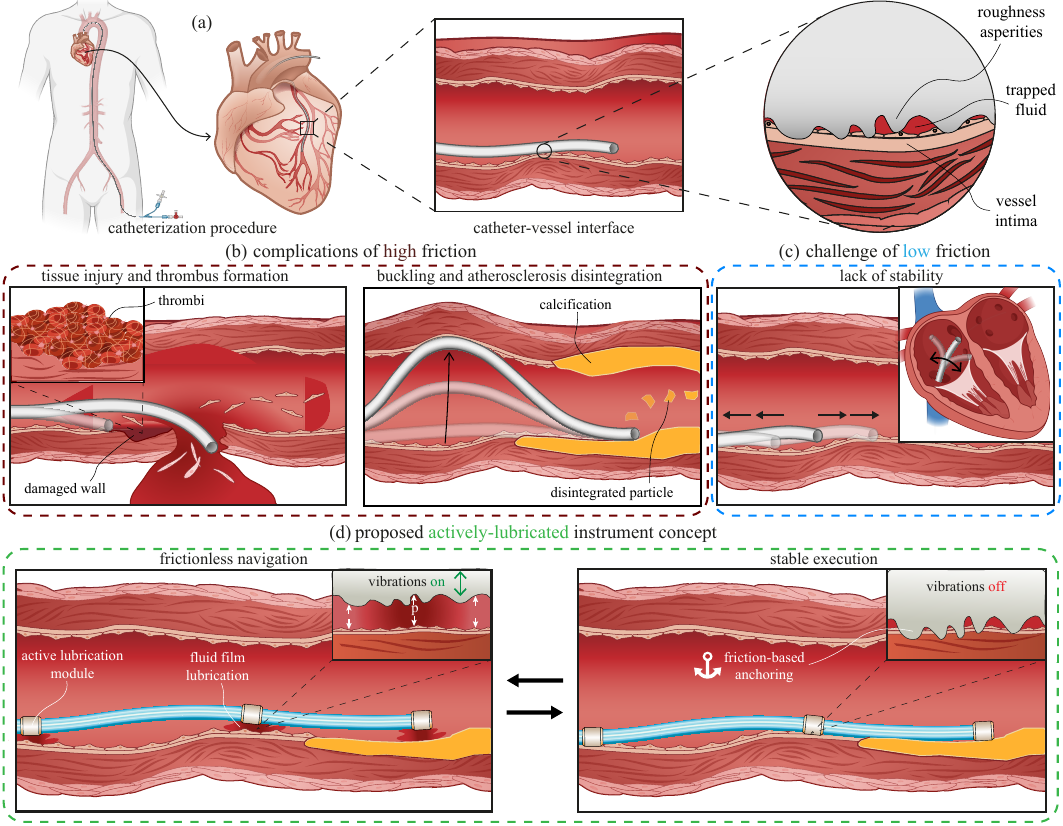}
    \caption{\textbf{Friction-induced complications in transluminal procedures and our proposed active lubrication approach.} (a) Catheter navigation through the femoral artery, with the inset illustrating asperity contact at the catheter--vessel interface. (b) Excessive friction can cause vessel injury, catheter buckling, and dislodgement of calcified deposits. (c) Insufficient friction reduces instrument stability and precision. (d) The active lubrication concept integrates discrete friction-control modules that generate ultrasonic vibrations, creating a pressurized fluid film at the interface for frictionless navigation. Deactivating the vibrations restores frictional contact, anchoring the catheter at the target site. Friction state is controlled simply by toggling the vibration on and off.}
    \label{fig:concept}
\end{figure*}
 
During navigation, the insertion of transluminal instruments, such as catheters and endoscopes, causes frictional interactions with surrounding tissue \cite{Wagner2021,Zhang2022}, as illustrated in Fig.~\ref{fig:concept}(a). This friction causes tissue damage, bleeding, and patient discomfort across catheterization \cite{Dellimore2013,Dellimore2014,Dellimore2016,Bostan2017,lin2020,wang2024friction}, colonoscopy \cite{Li2014,Waddingham2022,Kim2019}, and upper gastrointestinal endoscopy \cite{Lin2017,lin2019,lin2020lubricating}. It also diminishes haptic transparency and alters the interventionist's force perception, increasing the risk of accidental dissection or perforation \cite{Perreault2006,Okamura2010,VanderPutten2008}. Excessive friction can further cause flexible instruments to buckle (Fig.~\ref{fig:concept}b): the shaft accumulates elastic energy until suddenly released, potentially exerting intense forces that can cause catastrophic luminal damage \cite{Gopesh,mao2024} and, in robot-assisted procedures, further complicate instrument control \cite{Sakes2016,Baek2022}. In catheterization specifically, friction can damage vessel walls, promoting post-operative thrombus formation, and dislodge calcified particles that can subsequently obstruct critical blood vessels (Fig.~\ref{fig:concept}b).
 
However, while high friction is detrimental to the navigation of transluminal instruments, it becomes an asset when executing surgical operations such as cutting, puncturing, or suturing. All these operations require frictional anchoring between the instrument and the tissue in order to maintain position and apply force with precision~\cite{Muranishi2020,Loeve2010,Ranzani2017}. In low-friction conditions, particularly within open cavities such as the heart, the instability of the instrument may hinder effective manipulation and compromise procedural success (Fig.~\ref{fig:concept}c). This contradiction, where low friction is needed during navigation and high friction is needed when anchoring, is a significant design challenge for instruments, which must strike a balance between ease of navigation and stability.
 
Current practices rely on passive lubrication strategies to mitigate the risks of friction in transluminal interventions. To reduce friction during navigation, conventional strategies use liquid lubricants~\cite{Brocchi2008,Wilson2013Catheter,Hernandez2013,Wang2018,Watanabe2024} or self-lubricating coatings, such as hydrophilic coatings that become slippery when exposed to a wet surface~\cite{Wyman2012,Niemczyk2015}. However, these methods have notable limitations. Add-on lubricants are often short-lived, gradually losing effectiveness as they are wiped away during navigation \cite{Brocchi2008}. Hydrophilic coatings, though effective in specific procedures such as urology, lack versatility for broader transluminal applications. Crucially, neither approach offers dynamic control over friction—they cannot switch between low-friction and high-friction states based on the procedural phase. The inability of these techniques to meet the changing frictional demands of transluminal interventions highlights the need for active lubrication strategies that enable real-time, on-demand friction control for safer navigation and more stable procedure execution.
 
This paper introduces an active lubrication concept for transluminal instruments that enables real-time control of friction via ultrasonic lubrication. In this concept, we use high-frequency (ultrasonic) surface vibrations to create a pressurized squeeze film of fluid at the interface between the instrument and tissue, which actively lubricates the interface on demand, as illustrated in Fig.~\ref{fig:concept}(d). By dynamically switching friction states, the instrument can navigate safely to its destination and then transition to a stable configuration for surgical tasks. We present a versatile active lubrication sheath that modulates friction along the instrument shaft on demand, building upon our initial demonstration \cite{Atalla2024}. We show how effective active lubrication modules are at changing friction under varying mechanical loads, surface conditions, and anatomical curvatures. Finally, we validate the system in a cardiac catheterization use case using \textit{ex vivo} porcine aorta tissue, and assess its thermal safety during contact with biological tissues.

The remainder of this paper is organized as follows. Section II introduces the working principle and design of the active lubrication sheath. Section III describes the experimental methods used to evaluate its performance. Section IV presents the results and discusses their implications, and Section V concludes the paper.

\begin{figure*}
    \centering  \includegraphics[width=\textwidth]{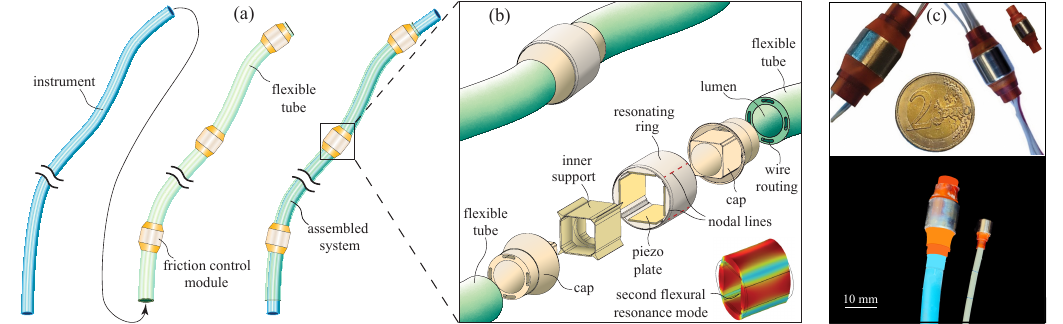}
    \caption{\textbf{Design concept and prototypes.} (a) The proposed active lubrication sheath featuring discrete friction control modules along its shaft. The instrument inserted into this sheath can be actively lubricated on demand. (b) Design of the friction control modules and resonance mode of the resonating ring. (c) Top: $15~\text{mm}$, $10~\text{mm}$, and $3~\text{mm}$ prototypes of the friction control modules. Bottom: fully assembled sheath with friction control module at the tip, $10~\text{mm}$ and $3~\text{mm}$ size.}
    \label{fig:design}
\end{figure*}
 
\section{Working Principle and Design of Active Lubrication Sheath}
\label{sec:design}
 
\subsection{Ultrasonic Lubrication}
When two surfaces are pressed together, contact occurs only at a finite number of microscopic roughness asperities. These asperities in contact form micro-junctions, which collectively constitute the real area of contact that carries the applied load. These micro-junctions govern sliding friction, as the deformation and shearing of these micro-junctions when relative sliding occurs dictates the friction force experienced by the two objects in contact.  The remaining regions of the interface, where microscopic asperities are not in contact, remain separated by small gaps, where fluid from the surrounding medium, whether air or liquid, is trapped. 

Ultrasonic lubrication exploits this trapped interfacial fluid. When transverse ultrasonic vibrations are applied to one of the contacting surfaces, the trapped fluid is repeatedly compressed and released. Consequently, the oscillatory motion generates a net positive pressure due to the non-linear response of the fluid to the periodic compression. This positive net pressure creates a fluid cushion between the two surfaces that partially levitates the vibrating surface away from the counter surface, decreasing asperity contact and thus reducing friction during sliding \cite{10.1115/1.3653080,Wiertlewski2016}.

The resulting friction reduction increases monotonically with vibration amplitude~\cite{Wiertlewski2016}, allowing friction to be controlled by adjusting the excitation amplitude. With the vibrations off, the interface remains in a high-friction contact state. When the vibrations are activated, the squeeze film supports part of the normal load and shifts the interface toward a low-friction state. This principle, commonly referred to as ultrasonic friction modulation or ultrasonic lubrication, has been applied in tactile displays~\cite{Winfield2007}, contactless manipulation~\cite{Gabai2019}, and squeeze-film bearings~\cite{Zhao2013,Shi2019}. We recently demonstrated that it also remains effective in liquid environments~\cite{Atalla2023}, making it particularly relevant to transluminal instruments operating against hydrated biological tissue.

\subsection{Conceptual Design of the Active Lubrication Sheath }
The active lubrication sheath features discrete friction control modules distributed along its length, enabling localized, on-demand friction control. When switched on, these modules generate controlled ultrasonic vibrations that induce fluid film lubrication at the module-tissue interface, which transforms the instrument into a low-friction mode for frictionless navigation, as illustrated in Fig.~\ref{fig:concept}(d). Once switched off, the modules re-establish contact with the luminal wall, creating friction-based anchoring for stable execution, as illustrated in Fig.~\ref{fig:concept}(d).

The sheath can function as a catheter or be used to transform existing instruments into actively lubricated tools by inserting them into the sheath, as illustrated in Fig.\ref{fig:design}(a). The modules are slightly larger in diameter than the sheath, to ensure that contact primarily occurs between the modules and the luminal walls, minimizing direct contact between the sheath body and tissue. Additionally, the sheath body is coated with an ultra-low friction material similar to \cite{mao2024} to ensure smooth interaction with the luminal wall, if contact occurs.

To effectively generate the desired active ultrasonic lubrication effect, the friction control modules should achieve vibration amplitudes of $\geq 2~\mathrm{\mu m}$ at an ultrasonic frequency $\geq 20~\mathrm{kHz}$~\cite{Biet,watanabe}.

\subsection{Detailed Design and Simulation}
Each friction control module generates ultrasonic transverse vibrations through a resonating ring structure equipped with four piezoelectric plates, as shown in Fig.\ref{fig:design}(b). When these plates are excited at the resonant frequency of the structure, the latter vibrates along its second flexural resonance mode, as depicted in Fig.\ref{fig:design}(b). The resonance frequency is determined by the diameter of the ring and wall thickness. The resonating ring is supported by a separable inner support structure made from a self-lubricating material, featuring needle-like edges that hold the ring at the nodal lines of its vibration mode, where the vibration amplitude is near zero. These needle-like supports give way to the nodal lines to rotate freely which minimizes the interference with the vibration. Additionally, the separable support facilitates the assembly of the piezoelectric plates, a crucial feature when miniaturizing the modules. To maintain an unobstructed instrument lumen, we incorporated specific wire-routing channels into the side cap and sheath wall~\cite{Abdelaziz2024}, allowing the power wires to be routed through them.

Three ring sizes, with diameters ranging from $3~\text{mm}$ to $15~\text{mm}$, were designed to operate in the second flexural resonance mode at a frequency above $20~\text{kHz}$ while generating vibration amplitudes exceeding $2~\upmu\text{m}$. Finite element analysis (FEA) guided the detailed ring design and provided estimates of its resonance characteristics. We first performed an eigenfrequency analysis to identify the resonance associated with the target second flexural mode. A subsequent frequency-response analysis evaluated the vibration amplitude and deformation pattern under harmonic excitation. By iteratively adjusting the ring dimensions, we obtained designs that met the frequency and amplitude requirements for ultrasonic lubrication. The $10~\text{mm}$ ring was selected for detailed simulation and subsequent experimental characterization, as it provides a practical balance between relevance to transluminal instruments and ease of fabrication and experimentation. For this design with titanium, the model predicted a second flexural resonance at $22.7~\text{kHz}$ and a peak vibration amplitude of approximately $4~\upmu\text{m}$ per $100~\text{V}$. 

\subsection{Prototype Development}
We implemented the designed ring structures in titanium using wire electrical discharge machining. Four $1~\text{W}$ ceramic piezoelectric plates (SMPL60W5T03R112, STEMiNC, Davenport, IA, USA), each $0.3~\text{mm}$ thick, were then cut to size and manually bonded around each ring using a custom wedge fixture that applied uniform pressure during curing. After bonding, $0.5~\text{mm}$ power wires were soldered to the plates and routed through the internal channels and side caps. Following this process, we fabricated three prototypes with diameters ranging from $3~\text{mm}$ to $15~\text{mm}$ to demonstrate the scalability of the design, as shown in Fig.~\ref{fig:design}(c). The $10~\text{mm}$ prototype was used for the subsequent characterization and validation experiments.

\section{Experimental Evaluation}
We experimentally evaluated the active lubrication system through three sets of experiments. First, we characterized and validated the vibration response of the fabricated modules. Second, we evaluated their active lubrication performance through a series of sliding friction experiments under varying loading, lubricity, and curvature conditions, including validation on \emph{ex vivo} tissue. Third, we assessed the thermal safety of the modules via thermal imaging experiments. Additionally, we conducted a catheter insertion experiment to demonstrate the practical effect of active lubrication on buckling.

\subsection{Vibration Characterization}
We characterized the vibration response of the fabricated modules using a single-point laser Doppler vibrometer (LDV) (OFV5000, Polytec GmbH, Germany). To access the full ring surface, the module was mounted on a servo motor that rotated it around its circumference, while a dual-axis galvanometer scanner (ScannerMAX Saturn 5B, Edmund Optics Inc., NJ, USA) steered the LDV beam along its length, following an approach similar to~\cite{Atalla2024}. A MATLAB interface synchronized the servo motor, scanner, LDV, and excitation signal through a National Instruments data acquisition card (USB-6356, National Instruments, TX, USA).

The characterization followed three steps. First, frequency sweeps at the centers of the four anti-nodal lines identified the resonant frequency. Second, a voltage sweep at resonance established the relationship between the input voltage and peak vibration amplitude. Third, a full-surface scan reconstructed the vibration mode and quantified the effective vibrating area. During this scan, the ring rotated in $5^{\circ}$ increments while the LDV recorded the vibration response at ten points along each longitudinal scan line. The effective vibrating area was then calculated as the portion of the ring surface vibrating at or above the target amplitude of $2~\upmu\mathrm{m}$.

\begin{figure}
\centering
\includegraphics[width=\linewidth]{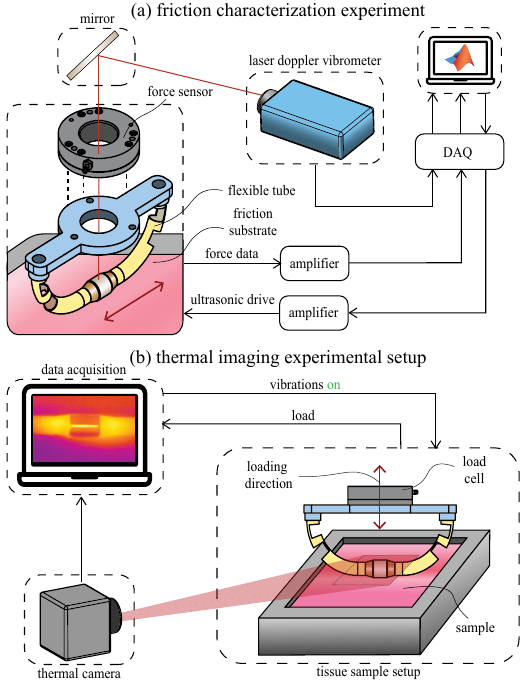}
\caption{\textbf{Experimental setups.} (a) Friction characterization setup combining force and vibration measurements during substrate sliding. (b) Thermal imaging setup for assessing thermal safety of ultrasonic lubrication.}
\label{fig:setups}
\end{figure}

\subsection{Active Lubrication Characterization}
We characterized the active lubrication performance using the sliding friction setup shown in Fig.~\ref{fig:setups}(a). The resonating ring was mounted on a six-axis force sensor (Nano43, ATI Industrial Automation, NC, USA), and the ring--sensor assembly was attached to a manual linear slider that controlled the normal load. A motorized linear stage moved the opposing substrate against the ring surface, while the force sensor recorded the friction force and the LDV measured the vibration amplitude from the top of the ring.

\begin{figure*}
    \centering
    \includegraphics[width=\linewidth]{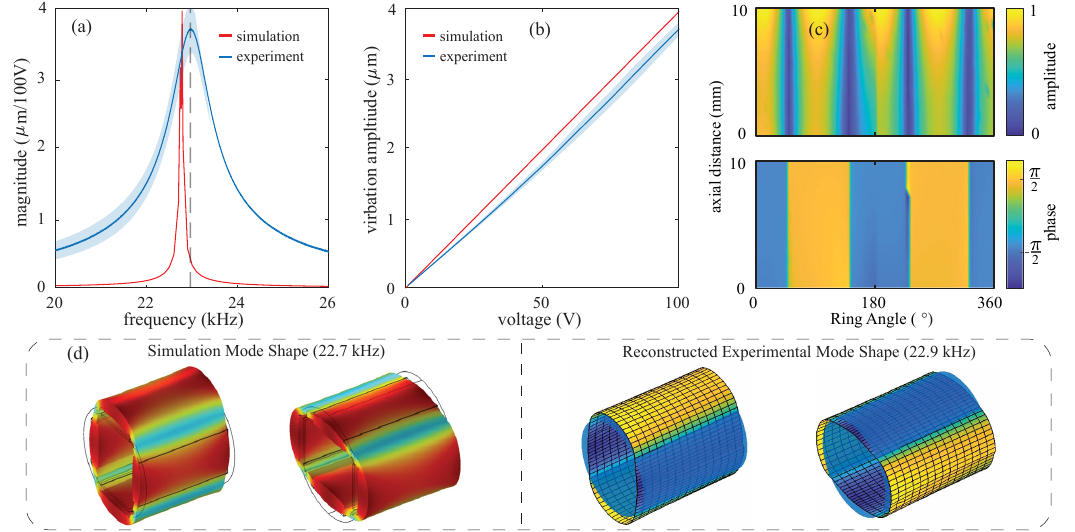}
    \caption{\textbf{Vibration characterization of the active lubrication modules.} (a) Frequency response of the resonating ring showing agreement between simulation and experiment. (b) The relationship between amplitude of vibration and input voltage applied to the piezoelectric actuators for simulated and experimental results (c) Mode shape of the resonating ring is the second flexural resonance mode matching simulation (d) comparison between the predicted mode shape by simulation and the experimentally reconstructed mode shape of the resonating ring prototype. (See Supplementary Video 1 for the video demonstration.)}
    \label{fig:vibCh}
\end{figure*}

The characterization followed two test series. In both, the substrate traveled $15~\mathrm{mm}$ at $1~\mathrm{mm/s}$ under the prescribed normal load, and each condition was repeated six times. First, the input voltage was increased linearly from its minimum to maximum value over $5~\mathrm{s}$ during sliding to establish the relationship between vibration amplitude and friction under different loading and surface lubricity conditions. Second, substrates with different curvatures were tested by activating the module at the maximum input voltage $5~\mathrm{s}$ after sliding began. The resulting friction reduction was then normalized by the corresponding vibration amplitude to quantify the friction reduction capacity (reduction percentage per unit amplitude).

\subsection{\emph{ex vivo} Tissue Validation}
We validated the friction-control modules on hydrated \emph{ex vivo} biological tissue using porcine aorta samples. Aortas from approximately six-month-old pigs were obtained from a local abattoir within $2~\mathrm{h}$ of slaughter and transported in phosphate-buffered saline (PBS) at $4^{\circ}\mathrm{C}$. After residual tissue was gently removed, rectangular sections approximately $10~\mathrm{mm}$ in size were cut along the arterial axis and stored at $-80^{\circ}\mathrm{C}$ for no longer than 30 days. Before testing, the samples were fully thawed and rehydrated in PBS.

Four tissue samples were tested under a normal load of $1~\mathrm{N}$. Each sample was cut open, flattened, and kept hydrated with PBS throughout the experiment, following~\cite{TAKASHIMA2007319,lin2020}. Six sliding trials were performed along each path. The first three trials were used to condition the tissue, while the final three were included in the analysis. Three distinct paths were tested on each sample to assess path dependence. During each trial, friction was first measured with the vibrations inactive and then immediately after activating the module, enabling a direct within-trial comparison of the coefficient of friction between the inactive and active states.

\subsection{Thermal Safety Assessment}

To assess thermal safety, heat generation at the module--tissue interface was monitored using a thermal camera (thermoIMAGER TIM~160S, Micro-Epsilon, NC, USA), as shown in Fig.~\ref{fig:setups}(b). The tissue was hydrated with PBS, and the module was loaded against its surface at $1~\mathrm{N}$. The module was then activated without sliding while the interface temperature was continuously recorded at an ambient temperature of $20$--$22^{\circ}\mathrm{C}$. Six tissue samples were tested, with three repetitions per sample.

\subsection{Catheter Buckling Demonstration}

Finally, we integrated the active lubrication module at the tip of a catheter to qualitatively demonstrate its effect on buckling during insertion into a 3D-printed lumen model. The catheter base was deliberately misaligned with the lumen axis to increase friction at the tip and create a challenging insertion condition. We compared insertions with and without activation to demonstrate how reducing friction at the catheter tip influences buckling and ease of insertion.

\section{Results and Discussion}
\subsection{Vibration Characterization}
The resonant frequency of the 10~mm prototype was measured at $22.9~\mathrm{kHz}$, with a linear vibration amplitude of slope $3.7~\upmu\mathrm{m}/100~\mathrm{V}$, in close agreement with the FEA predictions of $22.7~\mathrm{kHz}$ and $\approx 4~\upmu\mathrm{m}/100~\mathrm{V}$
(Fig.~\ref{fig:vibCh}(a)(b)). Surface scanning confirmed that the excited mode is the second flexural resonance mode (Fig.~\ref{fig:vibCh}(c)(d)), with $\approx 84\%$ of the ring surface vibrating at or above the $2~\upmu\mathrm{m}$ threshold required for effective ultrasonic lubrication. The minor amplitude discrepancy between simulation and experiment ($3.7$ vs $4~\upmu\mathrm{m}/100~\mathrm{V}$) can be attributed to fabrication tolerances and differences in material properties between model and prototype.

\begin{figure*}[t!]
    \centering  \includegraphics[width=\textwidth]{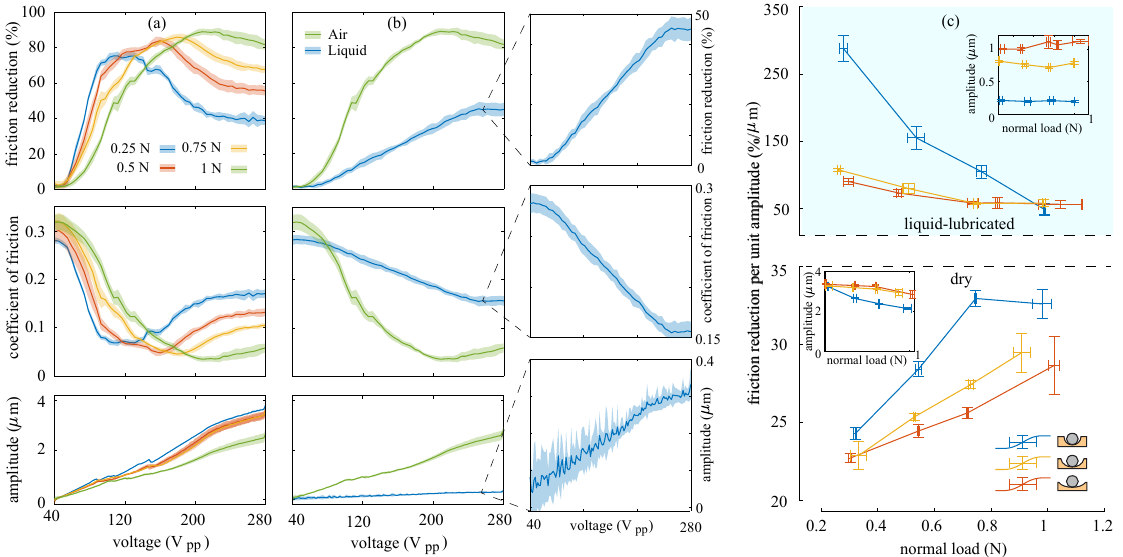}
    \caption{\textbf{Active lubrication performance under varying contact conditions.} (a) Friction reduction, coefficient of friction, and vibration amplitude as functions of input voltage at four normal loads. The vibration amplitude increases with voltage but decreases with increasing load, while friction reduction rises to a maximum before slightly declining at higher amplitudes. (b) Comparison between air and liquid environments, showing a more linear response and greater friction reduction for a given vibration amplitude in liquid. (c) Friction-reduction capacity across different normal loads and surface curvatures under dry and liquid-lubricated conditions. Under dry contact, the capacity increases with curvature and load, whereas under liquid lubrication it remains higher overall but decreases at larger loads. The insets show the corresponding vibration amplitudes. All conditions were repeated six times; lines and points represent the mean, while shaded regions and error bars indicate the standard deviation.}
    \label{fig:friction characterization}
\end{figure*}

\subsection{Friction Modulation Characterization}
The friction modulation experiment in dry contact revealed a linear correlation between the amplitude and the input voltage (Fig.~\ref{fig:friction characterization}(a)), consistent with the contact-free response in Fig.~\ref{fig:vibCh}(b), albeit with a shallower slope. As the load increases, the slope of the correlation shifts downwards due to the greater damping associated with higher loads. The corresponding friction reduction was robust to load, consistently achieving a similar maximum friction reduction in all loading conditions $\approx 80-82\%$, matching the maximum friction reduction achieved by ultrasonic haptic touchscreens \cite{Biet,Winfield2007}. However, increasing the normal load causes a horizontal offset in the friction reduction curve towards higher voltages. This offset is due to the necessity for higher vibration amplitudes to counteract the extra load, to maintain similar levels of friction reduction. Beyond the peak, friction reduction consistently declined across all loads, suggesting a saturation of the load-carrying capacity of the air squeeze film. Once the film pressure saturates, further energy input is converted to heat, increasing air viscosity, and consequently dissipation, which reduces the load-carrying capacity of the film and increases friction.

In liquid-lubricated contact, the amplitude-to-voltage mapping followed a consistently linear response but with an amplitude of $0.3~\upmu\mathrm{m}$ at maximum input voltage — approximately seven times lower than in air (Fig.~\ref{fig:friction characterization}(b)). Friction reduction was negligible below $\approx 80~\mathrm{V}$, below which amplitude was insufficient to overcome interfacial capillary adhesion, then followed a linear trend before saturating at $\approx 42\%$. While this maximum is roughly half the dry-contact value, the friction reduction per unit amplitude was significantly higher in liquid — consistent with the superior load-carrying capacity of incompressible liquid squeeze films predicted by theory~\cite{Atalla2023}. We also observe that liquid-lubricated contact exhibits a linear behavior for the full range of input voltage in comparison to dry contact. Importantly, in the liquid case, saturation of the reduction of friction is caused by a saturation of the vibration amplitude, unlike in the dry condition where the saturation comes from the load carrying capacity of the air film. 

\begin{figure*}[t!]
    \centering  \includegraphics[width=\textwidth]{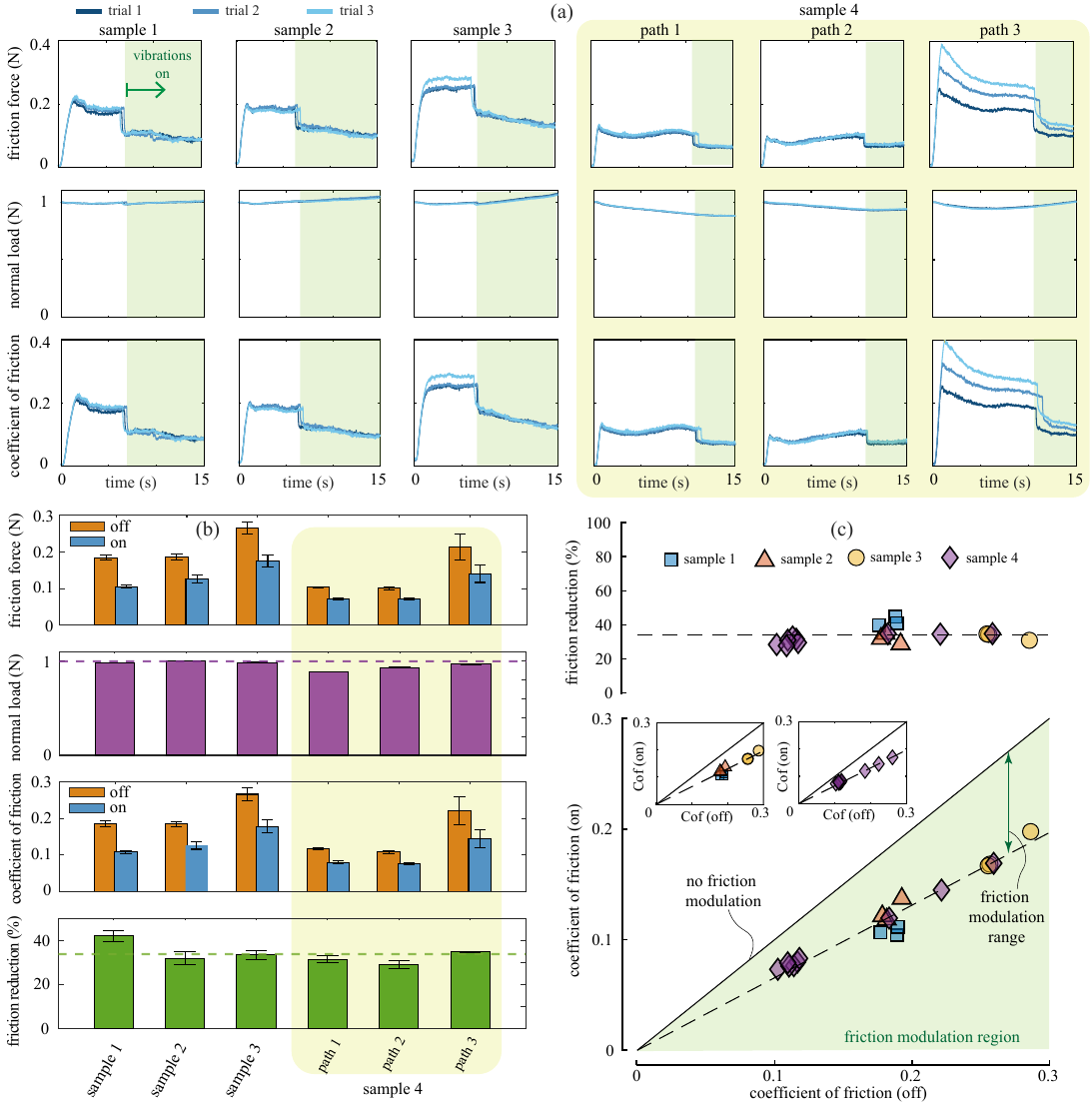}
     \caption{\textbf{Friction validation on \textit{ex vivo} porcine aorta tissue.} Each experiment was repeated six times; the first three repetitions conditioned the tissue and the last three were used for analysis. (a) Recorded friction force, normal load, and coefficient of friction across different samples and paths. Upon activating vibrations, friction drops immediately, demonstrating the active lubrication effect. (b) Comparison across samples and paths (mean $\pm$ standard deviation). Variability in friction and normal load is attributed to non-uniform tissue topography and partial drying of the samples over time. (c) Linear correlation between the coefficient of friction in the off condition $\mu_\mathrm{off}$ and the on condition $\mu_\mathrm{on}$.}
    \label{fig:ex-vivo}
\end{figure*}

\subsection{Effect of Surface Curvature}
In dry contact conditions, we observed that the vibration amplitude decreases with increasing load in all three curvature profiles, as shown in the inset of Fig.~\ref{fig:friction characterization}(c). This can be attributed to an increase in viscous losses caused by higher loads. Additionally, at a given load, the vibration amplitude decreases with increasing diameter ratio, which can be attributed to the larger contact area and the resulting increase in damping. For the friction reduction capacity, we found that at the same load, surfaces with higher curvature exhibited greater friction reduction capacity, which can be explained by the greater contact area generating a proportional increase in the squeeze film force \cite{Atalla2023}, as shown in Fig.~\ref{fig:friction characterization}(c). Moreover, the friction reduction capacity improved with increasing load across all curvature profiles. This trend is likely due to the reduced interfacial gap thickness between the two surfaces at higher loads, enhancing the squeeze film force, which results in greater friction reduction at the same vibration amplitude.

Looking at the liquid-lubricated contact in Fig.~\ref{fig:friction characterization}(c), we notice that the vibration amplitude, shown in the inset, remains almost constant across all loads. This constancy is likely due to the vibrating module being levitated and vibrating on top of a liquid squeeze film. In addition, the friction reduction capacity is reduced as the load increases, most likely because at higher loads, the interfacial gap thickness between the two surfaces becomes smaller, which in turn impedes the liquid from flowing in and out of the interface. This free movement of the liquid is a crucial factor for squeeze film pressure to develop in liquid films \cite{Atalla2023}. We observe that the most curved interface in the liquid-lubricated case initially exhibits significantly larger friction reduction capacity as compared to the other two profiles. It can be explained by the larger surface area contributing to an increased squeeze film force. However, the friction reduction capacity rapidly declines as the load increases due to the higher resistance for the liquid to move across the film. This larger resistance causes the liquid to lose more energy and thus reduces its load-carrying capacity. 

Based on the linear correlation, between friction and the vibration amplitude, we observed in the friction modulation experiment Fig.~\ref{fig:friction characterization}(b), one can assume that the behavior in the liquid-lubricated case, shown in Fig.~\ref{fig:friction characterization}(c) holds across different input voltages. However, for dry contact, the same conclusion may only be applied in the linear region of the response in Fig.~\ref{fig:friction characterization}(a), while the other region is likely voltage-dependent. 

\begin{figure}[t!]
    \centering  \includegraphics[width=\linewidth]{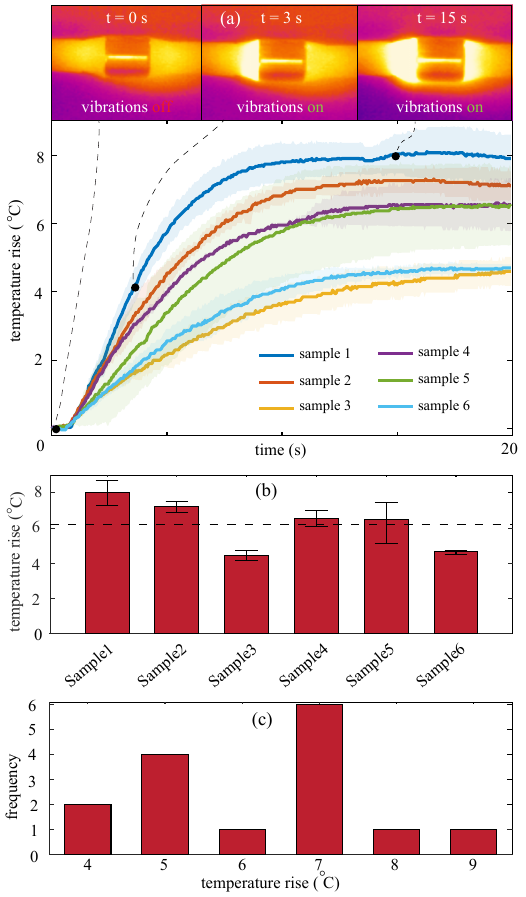}
    \caption{\textbf{Thermal safety assessment results.} Each experiment was repeated three times; solid lines and bars represent the mean, shadings and error bars the standard deviation. (a) Time response of the temperature rise across six tissue samples with representative thermal images. Temperature increases upon activation and reaches steady state within approximately $10~\text{s}$. (b) Steady-state temperature rise per sample, averaging $6.2^{\circ}\text{C}$. (c) Histogram of temperature rise, with $7^{\circ}\text{C}$ being the most frequent value, within the established thermal safety limit \cite{pavel,Nu}. (See Supplementary Video 2.)}
    \label{fig:thermalSafety}
\end{figure}

We conclude that highly curved surfaces are more favorable for friction reduction in both dry and liquid-lubricated environments. This conclusion is supported by the fact that friction force does not depend on the apparent area of contact. As curvature increases, the apparent area of contact increases, enhancing the squeeze-film force without affecting the friction force, resulting in a net increase in the friction reduction capacity.

\begin{figure*}
    \centering \includegraphics[width=\textwidth]{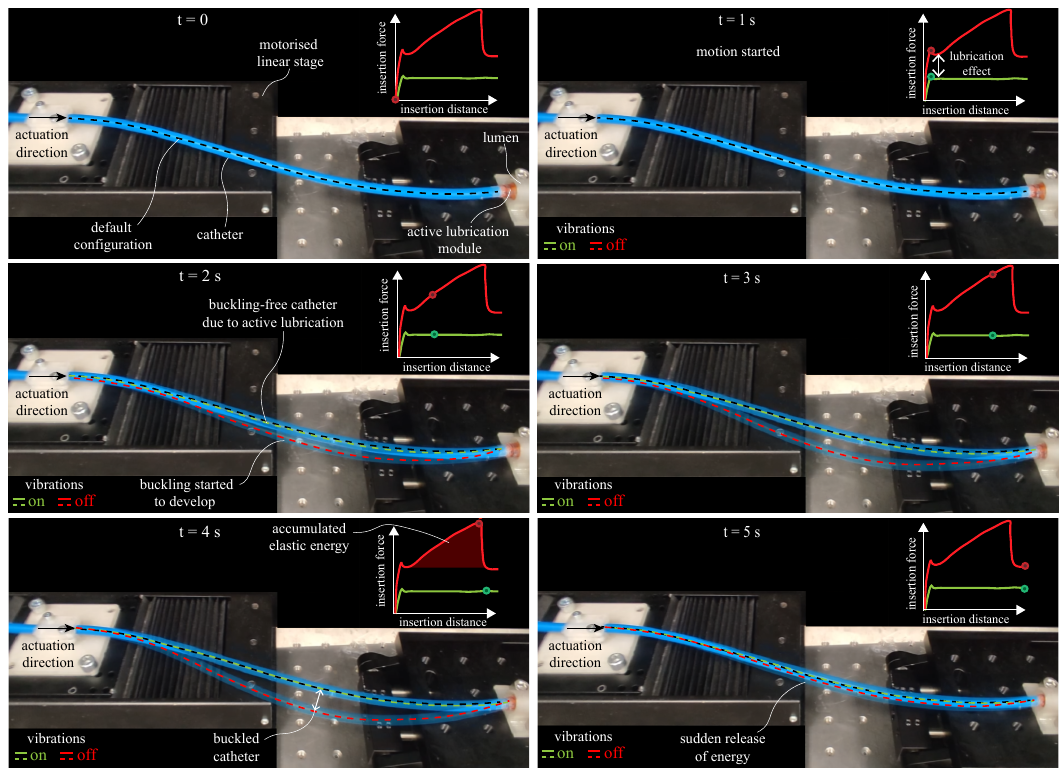}
    \caption{\textbf{Demonstration of active lubrication preventing catheter buckling during insertion.} A catheter with a friction-control module at its tip was inserted into a 3D-printed lumen, with the base deliberately misaligned to promote frictional wall interaction. Without lubrication, the catheter buckled, accumulating elastic energy in the shaft until it was suddenly released, potentially exerting excessive shock loads on the lumen wall. With lubrication activated, the catheter navigated smoothly without buckling. (See Supplementary Video 3.)}
    \label{fig:bucklingDemo}
\end{figure*}

\subsection{\textit{Ex Vivo} Tissue Validation}
Upon enabling vibration, the coefficient of friction dropped immediately to a lower level as shown in Fig.~\ref{fig:ex-vivo}(a), achieving friction reduction up to $42\%$ and an average of $38\%$ across all tissue samples and paths at maximum vibration amplitude as shown in Fig.~\ref{fig:ex-vivo}(b). 

Looking closely at the results in Fig.~\ref{fig:ex-vivo}(a)\&(b), we first observe that normal load varies over time across the different samples and paths with small variance across trials. These variations result from the non-uniform topography of the tissue samples along the sliding path. Second, we observe that the starting coefficient of friction (before vibrations are activated) $\mu_\mathrm{off}$ varied between samples, paths and between trials (example in Fig.~\ref{fig:ex-vivo}(a) - Sample 4-Path 3). This variation across samples and paths can be attributed to the uneven surface topography of tissue samples which resulted in uneven distribution of the surface lubrication layer. In addition, we observed that the tissue samples were drying out over time, resulting in increasing friction levels across the different trials. 

This seemingly undesired variance in the coefficient of friction revealed an interesting correlation between the coefficient of friction in the off condition (before vibrations are activated) $\mu_\mathrm{off}$ and the coefficient of friction in the on condition (after vibrations are activated) $\mu_\mathrm{on}$ as shown in Fig.~\ref{fig:ex-vivo}(c). The coefficient of friction in the on condition $\mu_\mathrm{on}$ is positively correlated to the coefficient of friction in the off condition $\mu_\mathrm{off}$ and this correlation holds for both different samples and different paths as shown in the insets of Fig.~\ref{fig:ex-vivo}(c). This correlation is well explained by the model of Wiertlewski et al. \cite{Wiertlewski2016} in which they showed using multi-scale contact theory~\cite{greenwood1966contact,BUSH197587,persson} and the squeeze film theory \cite{langlois1962isothermal,10.1115/1.3653080} that the coefficient of friction after enabling vibrations $\mu_\mathrm{on}$ is proportional to the coefficient of friction without vibrations $\mu_\mathrm{off}$ as follows:
\begin{equation}
    \mu_\mathrm{on} = \beta. \mu_\mathrm{off}.exp(\frac{\alpha}{N})
\end{equation}

where $\alpha$ is the vibration amplitude, $N$ is the normal load and $\beta$ is a proportionality constant that depends on surface roughness and material elasticity. This equation implies that the slope of the friction modulation line depends on the vibration amplitude and normal load. At identical vibration amplitude and load, the range between $\mu_\mathrm{off}$ and $\mu_\mathrm{on}$, as shown in Fig.~\ref{fig:ex-vivo}(c), increases linearly with the increase in $\mu_\mathrm{off}$, while the friction reduction percentage (\%) remains constant. 

This correlation between the coefficient of friction in on ($\mu_\mathrm{on}$) and off ($\mu_\mathrm{off}$) conditions, shown in Fig.~\ref{fig:ex-vivo}(c), provides key design insights for optimizing the active lubrication system. Specifically, it is beneficial to maximize both the starting coefficient of friction ($\mu_\mathrm{off}$) and the vibration amplitude. Higher $\mu_\mathrm{off}$ increases the difference between $\mu_\mathrm{off}$ and $\mu_\mathrm{on}$ at maximum vibration amplitude, enabling more effective switching between friction states. Similarly, maximizing vibration amplitude increases the controllable range between $\mu_\mathrm{off}$ and $\mu_\mathrm{on}$ for a given starting coefficient of friction $\mu_\mathrm{off}$. Combining both factors can potentially result in a wider controllable range, with higher friction in the "off" state and lower friction in the "on" state. However, further investigation is needed to determine the optimal controllable range and the acceptable limits for interacting with tissues. That is to find the most effective range of controllable friction states ($\mu_\mathrm{off}$ and $\mu_\mathrm{on}$) that minimizes tissue damage while ensuring stability during catheterization.

\subsection{Thermal Safety}
Upon activation, the module--tissue interface temperature rose immediately and reached a steady state after approximately ten seconds, as shown in Fig.~\ref{fig:thermalSafety}(a). We found the average temperature rise to be $\approx 6.2^{\circ}\mathrm{C}$, as shown in Fig.~\ref{fig:thermalSafety}(b), with $7^{\circ}\mathrm{C}$ being the most frequent temperature rise as shown in Fig.~\ref{fig:thermalSafety}(c). This temperature rise translates to $\approx 44^{\circ}\mathrm{C}$ inside the human body which is within the thermal safety range for the human body \cite{pavel,Nu}, suggesting the potential thermal safety of this technology for transluminal surgery applications. 


\subsection{Catheter Buckling Demonstration}
As shown in Fig.~\ref{fig:bucklingDemo}, without lubrication, the catheter tip experienced higher friction, causing the catheter to buckle. This buckling resulted in elastic energy accumulation in the shaft of the catheter until it was suddenly released; rapidly advancing the catheter tip forward. This sudden release of energy can exert excessive forces on the luminal wall, potentially causing catastrophic lumen dissection or rupture \cite{Gopesh, mao2024}. In contrast, with the lubrication module activated, the catheter tip moved smoothly through the lumen without buckling, as seen in Fig.~\ref{fig:bucklingDemo}, maintaining low frictional interaction with the lumen wall. This demonstration underscores the positive impact of the active lubrication system in reducing catheter buckling, potentially enhancing insertion and navigation, and thus ultimately improving the safety and efficacy of transluminal procedures.

\subsection{Limitations and Future Work}
The current study has several limitations that should be considered when interpreting the results. Friction was characterized on rigid 3D-printed substrates and one soft tissue type, porcine aorta. Although these represent calcified and healthy vascular tissue, respectively, they do not cover the full range of tissue properties encountered in transluminal procedures. Thermal safety was also tested under static and dry conditions at room temperature, without sliding or liquid flow. The measured temperature rise may therefore differ from that at body temperature and likely overestimates the heating expected during in vivo use, where blood flow would remove heat through convection. In addition, the contact point between the module and the tissue would continuously shift during catheter navigation, limiting the duration of local heat exposure. These effects need to be examined under more physiologically realistic conditions. The current prototype also lacks waterproofing, which prevents fully submerged operation, while the spacing, diameter, and number of modules along the sheath have not yet been optimized. Finally, the effects of prolonged ultrasonic vibration on tissue have not yet been studied at the histological level.

Future work will extend the validation to additional tissue types, including intestinal tissue, to assess how well the approach transfers to endoscopic and other transluminal procedures. Histological analysis will also be used to determine whether prolonged exposure to the vibrations causes tissue damage. A waterproof module will be developed to support fully submerged testing, followed by in vivo animal studies using complete instrument assemblies. The spacing, diameter, and number of modules along the sheath will also be refined to improve friction control without reducing instrument flexibility. In parallel, thermal testing will be repeated at body temperature and under flow or perfusion to better represent physiological conditions and define a more reliable safety baseline.

The same principle may also be useful beyond cardiovascular applications in procedures where friction affects instrument navigation, such as flexible endoscopy and bronchoscopy. The characterization across normal load, lubrication, and surface curvature provides a basis for adapting the module design to these different anatomical and procedural conditions.

\section{Conclusion}
This paper presented an active lubrication sheath for transluminal medical instruments that controls friction on demand through discrete ultrasonic modules distributed along the shaft. The modules operated at $22.9~\mathrm{kHz}$, reached a peak vibration amplitude of $3.7~\upmu\mathrm{m}/100~\mathrm{V}$, and provided effective lubrication over approximately 84\% of their surface. Tribometric experiments showed friction reductions of up to 82\% on rigid surfaces and 42\% on PBS-lubricated \emph{ex vivo} porcine aorta tissue, representing simplified calcified and healthy vascular interfaces, respectively. The results also showed a clear relation between ($\mu_\mathrm{off}$) and ($\mu_\mathrm{on}$), indicating that a wider controllable friction range requires both a higher initial coefficient of friction and a larger vibration amplitude. Active lubrication was most effective under liquid-lubricated conditions and generally improved with surface curvature, possibly because of the larger contact area. The average temperature rise was $6.2^{\circ}\mathrm{C}$ under static and dry test conditions, although physiological testing is still needed to assess the thermal response during use. In the catheter insertion experiment, the reduced wall friction lowered the tendency to buckle and enabled smoother advancement through the lumen. These findings support the use of ultrasonic lubrication in cardiovascular instruments and other transluminal devices, while future work should focus on waterproof modules, physiological thermal testing, additional tissue types, histological assessment, and optimization of the module arrangement along the shaft.

\section*{Acknowledgment}
This work was funded by a Delft University of Technology cohesion grant. The experimental work was supported by an EPSRC (UK) Centre-to-Centre Grant ``Tribology as an Enabling Technology (TRENT)'' (EP/S030476/1).

\section*{Data Availability}
All relevant data are available from the corresponding authors upon request.

\bibliographystyle{IEEEtran}
\bibliography{sn-bibliography}

\end{document}